\documentstyle[12pt]{article}

\begin{document}
\baselineskip=20.5pt
\def\beqra{\begin{eqnarray}} \def\eeqra{\end{eqnarray}}
\def\beqast{\begin{eqnarray*}} \def\eeqast{\end{eqnarray*}}
\def\beq{\begin{equation}}      \def\eeq{\end{equation}}
\def\be{\begin{enumerate}}   \def\ee{\end{enumerate}}

\def\fnote#1#2{\begingroup\def\thefootnote{#1}\footnote{#2}\addtocounter
{footnote}{-1}\endgroup}



\def\gam{\gamma}
\def\Gam{\Gamma}
\def\la{\lambda}
\def\eps{\epsilon}
\def\La{\Lambda}
\def\si{\sigma}
\def\Si{\Sigma}
\def\al{\alpha}
\def\Th{\Theta}
\def\th{\theta}
\def\tnu{\tilde\nu}
\def\vphi{\varphi}
\def\del{\delta}
\def\Del{\Delta}
\def\ab{\alpha\beta}
\def\om{\omega}
\def\Om{\Omega}
\def\mn{\mu\nu}
\def\mun{^{\mu}{}_{\nu}}
\def\kap{\kappa}
\def\rsi{\rho\sigma}
\def\beal{\beta\alpha}
\def\til{\tilde}
\def\rta{\rightarrow}
\def\eqv{\equiv}
\def\nab{\nabla}
\def\pa{\partial}
\def\sit{\tilde\sigma}
\def\ul{\underline}
\def\indt{\parindent2.5em}
\def\nd{\noindent}
\def\rsi{\rho\sigma}
\def\beal{\beta\alpha}
\def\caa{{\cal A}}
\def\cb{{\cal B}}
\def\cac{{\cal C}}
\def\cd{{\cal D}}
\def\ce{{\cal E}}
\def\cf{{\cal F}}
\def\cg{{\cal G}}
\def\cah{{\cal H}}
\def\ci{{\cal I}}
\def\cj{{\cal{J}}}
\def\ck{{\cal K}}
\def\cl{{\cal L}}
\def\cm{{\cal M}}
\def\cn{{\cal N}}
\def\cO{{\cal O}}
\def\cp{{\cal P}}
\def\car{{\cal R}}
\def\cs{{\cal S}}
\def\ct{{\cal{T}}}
\def\cu{{\cal{U}}}
\def\cv{{\cal{V}}}
\def\cw{{\cal{W}}}
\def\cx{{\cal{X}}}
\def\cy{{\cal{Y}}}
\def\cz{{\cal{Z}}}
\def\asymptotic{{_{\stackrel{\displaystyle\longrightarrow}
{x\rightarrow\infty}}\,\, }} 
\def\asymptext{\raisebox{.6ex}{${_{\stackrel{\displaystyle\longrightarrow}
{x\rightarrow\pm\infty}}\,\, }$}} 
\def\epsilim{{_{\textstyle{\rm lim}}\atop
_{~~~\epsilon\rightarrow 0+}\,\, }} 
\def\Llim{{_{\textstyle{\rm lim}}\atop
_{~~~L\rightarrow \infty}\,\, }} 
\def\omegalim{{_{\textstyle{\rm lim}}\atop
_{~~~\omega^2\rightarrow 0+}\,\, }} 
\def\xlimp{{_{\textstyle{\rm lim}}\atop
_{~~x\rightarrow \infty}\,\, }} 
\def\xlimm{{_{\textstyle{\rm lim}}\atop
_{~~~x\rightarrow -\infty}\,\, }} 
\def\asymptoticp{{_{\stackrel{\displaystyle\longrightarrow}
{x\rightarrow +\infty}}\,\, }} 
\def\asymptoticm{{_{\stackrel{\displaystyle\longrightarrow}
{x\rightarrow -\infty}}\,\, }} 

\def\notxi{\ \hbox{{$x_i$}\kern-.61em\hbox{\raisenot}}}
\def\notxj{\ \hbox{{$x_j$}\kern-.61em\hbox{\raisenot}}}
\def\notxk{\ \hbox{{$x_k$}\kern-.61em\hbox{\raisenot}}}
\def\notyi{\ \hbox{{$y_i$}\kern-.61em\hbox{\raisenot}}}
\def\notyj{\ \hbox{{$y_j$}\kern-.61em\hbox{\raisenot}}}
\def\notyk{\ \hbox{{$y_k$}\kern-.61em\hbox{\raisenot}}}
\def\raisenot{\raise .5mm\hbox{/}}
\def\nota{\ \hbox{{$a$}\kern-.49em\hbox{/}}}
\def\notA{\hbox{{$A$}\kern-.54em\hbox{\raisenot}}}
\def\notb{\ \hbox{{$b$}\kern-.47em\hbox{/}}}
\def\notB{\ \hbox{{$B$}\kern-.60em\hbox{\raisenot}}}
\def\notc{\ \hbox{{$c$}\kern-.45em\hbox{/}}}
\def\notd{\ \hbox{{$d$}\kern-.53em\hbox{/}}}
\def\notbd{\ \hbox{{$D$}\kern-.61em\hbox{\raisenot}}} 
\def\note{\ \hbox{{$e$}\kern-.47em\hbox{/}}}
\def\notk{\ \hbox{{$k$}\kern-.51em\hbox{/}}}
\def\notp{\ \hbox{{$p$}\kern-.43em\hbox{/}}}
\def\notq{\ \hbox{{$q$}\kern-.47em\hbox{/}}}
\def\notW{\ \hbox{{$W$}\kern-.75em\hbox{\raisenot}}}
\def\notz{\ \hbox{{$Z$}\kern-.61em\hbox{\raisenot}}}
\def\notpa{\hbox{{$\partial$}\kern-.54em\hbox{\raisenot}}}
\def\fo{\hbox{{1}\kern-.25em\hbox{l}}}  
\def\rf#1{$^{#1}$}
\def\bx{\Box}
\def\tr{{\rm Tr}}
\def\rmtr{{\rm tr}}
\def\dgg{\dagger}
\def\lag{\langle}
\def\rag{\rangle}
\def\bmid{\big|}
\def\vlap{\overrightarrow{\La p}} 
\def\lrta{\longrightarrow} \def\lrar{\raisebox{.8ex}{$\longrightarrow$}}
\def\ON{{\cal O}(N)}
\def\UN{{\cal U}(N)}
\def\bdPh{\mbox{\boldmath{$\dot{\!\Phi}$}}}
\def\bPh{\mbox{\boldmath{$\Phi$}}}
\def\bPhs{\bPh^2}
\def\sef{S_{eff}[\sigma,\pi]}
\def\sigx{\sigma(x)}
\def\pix{\pi(x)}
\def\bph{\mbox{\boldmath{$\phi$}}}
\def\bphs{\bph^2}
\def\ex{\BM{x}}
\def\exs{\ex^2}
\def\xdot{\dot{\!\ex}}
\def\y{\BM{y}}
\def\ys{\y^2}
\def\ydot{\dot{\!\y}}
\def\pat{\pa_t}
\def\pax{\pa_x}
\def\hp{{\pi\over 2}}

\renewcommand{\thesection}{\arabic{section}}
\renewcommand{\theequation}{\thesection.\arabic{equation}}

\begin{flushright}
\today\\
\end{flushright}

\vspace*{.1in}
\begin{center}
  \Large{\sc Fredholm's Minors of Arbitrary Order: Their Representations 
as a Determinant of Resolvents and in Terms of Free Fermions and an 
Explicit Formula for Their Functional Derivative}
\normalsize

\vspace{36pt}
{\large Joshua Feinberg\fnote{*}{{\it e-mail: joshua@physics.technion.ac.il}}}
\\
\vspace{12pt}
{\small \em Department of Physics,}\\ 
{\small \em Oranim-University of Haifa, Tivon 36006, Israel}\fnote{**}
{permanent address}\\
{\small and}\\
{\small \em Department of Physics,}\\
{\small \em Technion - Israel Institute of Technology, Haifa 32000 Israel}

\vspace{.6cm}

\end{center}

\begin{minipage}{5.8in}
{\abstract~~~We study the Fredholm minors associated with a Fredholm 
equation of the second type. We present a couple of new linear recursion 
relations involving the $n$th and $n-1$th minors, whose solution is a 
representation of the $n$th minor as an $n\times n$ determinant of 
resolvents. The latter is given a simple interpretation in terms of a 
path integral over non-interacting fermions. We also provide an explicit 
formula for the functional derivative of a Fredholm minor of order $n$ with 
respect to the kernel. Our formula is a linear combination of the $n$th and 
the $n\pm 1$th minors. }

\end{minipage}

\vspace{58pt}

\vfill
\pagebreak

\setcounter{page}{1}

\section{Introduction}
\setcounter{equation}{0}

The ubiquity of linear integral equations, and in particular of Fredholm 
equations (FE) \cite{hilbert,tricomi,pog}, in mathematical physics, and more 
broadly in analysis, cannot be overstated. Thus, new results in this 
classical field, such as the results presented in this paper 
(Eqs.(\ref{determinantalrep}) and (\ref{derivative})), should be of 
some interest. 

To get oriented, let us briefly recall the basic definitions and facts
of Fredholm theory, relevant to our discussion. (We have adopted throughout 
this paper the conventions and notations of Chapter II of \cite{pog}.) 
Thus, consider a Fredholm integral equation of the second type in the 
unknown function $\phi(x)$, 
\beq\label{fredholm}
\phi(x) = f(x) + \lambda\int\limits_\Om N(x,y)\phi(y)\,dy\, ,
\eeq
with kernel $N(x,y)$ and given function $f(x)$. For simplicity, we shall
take $N(x,y)$ and $f(x)$ as real functions. The generalization to the complex 
case is straightforward. The complex variable $\lambda$ is the spectral 
parameter of the equation, and $\Om$ is the domain on which the equation is 
defined. To be concrete, we shall take $\Om$ as a compact domain of 
$N$-dimensional Euclidean space of volume $V$ 
\beq\label{volume}
\int\limits_\Om\,dx = V\,.
\eeq
We further assume that $N(x,y)$ is bounded on $\Om$
\beq\label{bounded}
|N(x,y)|\leq M\,,
\eeq
and that it is integrable in $\Om$ with respect to both $x$ and $y$. The 
given function $f(x)$ is assumed integrable as well. 

It is also useful to introduce the operator $\hat N$ and the vectors 
$ |\phi\rangle$ and $ |f\rangle$, which correspond to the kernel $N(x,y)$ 
and functions $\phi(x)$ and $f(x)$. Thus, in obvious notations,  
\beq\label{operatorial}
N(x,y) = \langle x | \hat N |y \rangle,\, f(x) = \langle x  |f\rangle\,,
\quad {\rm and}\quad \phi(x)=  \langle x |\phi\rangle\,. 
\eeq
In terms of (\ref{operatorial}), we can write Fredholm's equation
(\ref{fredholm}) as 
\beq\label{fredholmop}
\left({\bf 1} -\lambda\hat N\right) |\phi\rangle = |f\rangle\,.
\eeq

Next, we define the $n\times n$ determinant
\beqra\label{determinant}
N\left(\begin{array}{cccc} x_1 & x_2 & \dots & x_n \\
{}&{}&{}&{}\\ y_1 & y_2 & \dots & y_n 
\end{array}\right) = \det_{i,j} N(x_i,y_j)\,,
\eeqra
where $x_1,\ldots y_n$ is a set of $2n$ points in $\Om$. We shall refer to 
the $x_i$ as the row indices, and to the $y_j$ as the column indices of the
symbol on the left-hand side of (\ref{determinant}).
In some of the mathematical literature (\ref{determinant}) is known as the 
Fredholm determinant, however, we shall reserve this name, as is customary in 
vast portions of the physics and mathematics literature, to Fredholm's first 
series $D(\lambda)$ defined below in (\ref{firstseries}). 

Given (\ref{bounded}), it follows from a theorem due to Hadamard that 
(\ref{determinant}) is bounded according to 
\beq\label{hadamard}
\Bigg| N\left(\begin{array}{cccc} x_1 & x_2 & \dots & x_n \\
{}&{}&{}&{}\\ y_1 & y_2 & \dots & y_n 
\end{array}\right)\Bigg| \leq n^{n\over 2} M^n\,.
\eeq

Fredholm's $n$th minor is defined by the series
\beqra\label{nminor}
&&D_n\left(\begin{array}{cccc} x_1 & x_2 & \dots & x_n \\
{}&{}&{}&{}\\ y_1 & y_2 & \dots & y_n 
\end{array}\Bigg|\lambda\right) = 
N\left(\begin{array}{cccc} x_1 & x_2 & \dots & x_n \\
{}&{}&{}&{}\\ y_1 & y_2 & \dots & y_n 
\end{array}\right)\nonumber\\{}\nonumber\\{}\nonumber\\
&&+\sum_{p=1}^\infty\, {(-\lambda)^p\over p!}
\int\limits_\Om\,N\left(\begin{array}{cccccccc} x_1 & x_2 & \dots & x_n &
 s_1 & s_2 & \dots & s_p \\
{}&{}&{}&{}&{}&{}&{}&{}\\ y_1 & y_2 & \dots & y_n &
 s_1 & s_2 & \dots & s_p  \end{array}\right)\,ds_1\ldots ds_p\nonumber\\
\eeqra
By definition, $D_n$ is completely antisymmetric in the $x_i$, and also 
in the $y_i$. 
In view of (\ref{volume}) and (\ref{hadamard}), it is easy to see that 
the series (\ref{nminor}) converges absolutely to an entire function 
of $\lambda$.  

Fredholm's first series
\beqra\label{firstseries}
&&D (\lambda)  = 1 +\sum_{p=1}^\infty\, {(-\lambda)^p\over p!}
\int\limits_\Om\,N\left(\begin{array}{cccc} s_1 & s_2 & \dots & s_p \\
{}&{}&{}&{}\\ s_1 & s_2 & \dots & s_p  \end{array}\right)
\,ds_1\ldots ds_p
\eeqra
and second series 
\beqra\label{secondseries}
&&D (x,y; \lambda) = 
N(x,y) +\sum_{p=1}^\infty\, {(-\lambda)^p\over p!}
\int\limits_\Om\,N\left(\begin{array}{ccccc} x & s_1 & s_2 & \dots & s_p \\
{}&{}&{}&{}&{}\\ y & s_1 & s_2 & \dots & s_p  \end{array}\right)
\,ds_1\ldots ds_p\nonumber\\
\eeqra
correspond to setting $n=0$ and $n=1$ in (\ref{nminor}), respectively.

Fredholm's first series (\ref{firstseries}) is, by construction, the 
functional determinant 
\beq\label{fredholmdeterminant}
D(\lambda) = {\rm Det}\,\left({\bf 1} -\lambda\hat N\right) 
\eeq
of the operator on the left-hand side of (\ref{fredholmop}). It is usually 
known in the literature as the {\em Fredholm determinant} associated with 
(\ref{fredholm}), and we shall adhere to this convention here.

From the definitions (\ref{nminor}) and (\ref{firstseries}) we can prove
the important relation 
\beq\label{nthderivative}
{d^n D (\lambda)\over d\lambda^n} = (-1)^n\int\limits_\Om\,
D_n\left(\begin{array}{cccc} x_1 & x_2 & \dots & x_n \\
{}&{}&{}&{}\\ x_1 & x_2 & \dots & x_n \end{array}\Bigg|\lambda\right)
\,dx_1\ldots dx_n
\eeq
in a straightforward manner.

The motivation for introducing the minors stems from their important roles in
solving Fredholm's equation (\ref{fredholm}) in the most general case. This is
briefly reviewed in the appendix. In particular, for values of $\lambda$ 
such that $D(\lambda)\neq 0$, the solution of (\ref{fredholm}) is determined
by the resolvent kernel $R(x,y;\lambda)$ according to (\ref{uniquesolution}). 
$R(x,y;\lambda)$, in turn, is given in (\ref{resolvent}) as 
${D (x,y; \lambda)\over D(\lambda)}$. For values of $\lambda$ such that 
$D(\lambda)= 0$, the solution is given by (\ref{Phii}) and (\ref{particular}),
and involves the higher minors. 

The rest of the paper is organized as follows. In the next section we derive
our new recursion relations (Eqs.(\ref{yirecursionrelation1}) and 
(\ref{xirecursionrelation})) for the minors (\ref{nminor}). We then solve it
and obtain the representation (\ref{determinantalrep}) for (\ref{nminor}) 
as an $n\times n$ determinant over resolvents. In section 3 we provide an 
interpretation of this representation in terms of non-interacting
fermions. Finally, in section 4 we provide an explicit formula 
(Eq.(\ref{derivative})) for the functional derivative of the $n$th minor with 
respect to the kernel.

\section{The Minor $D_n$ as an $n\times n$ Determinant}
\setcounter{equation}{0}
\renewcommand{\theequation}{2.\arabic{equation}}

We start by deriving a couple of integral equations satisfied by 
$D_n$\,\cite{pog}. To obtain the first equation, expand each of the 
determinants in the series (\ref{nminor}) with respect to the row $x_i$, and 
integrate with respect to all the $s$-variables which occur in that term. 
It is easy to see, by elementary permutations of rows and columns, that in 
the $p$th term, all columns $s_1,\ldots, s_p$ yield the same integrated 
contribution. Then, after resumming over $p$, one obtains 
\beqra\label{xirow}
&&D_n\left(\begin{array}{ccc} x_1 & \dots & x_n \\
{}&{}&{}\\ y_1 & \dots & y_n 
\end{array}\Bigg|\lambda\right) =\nonumber\\{}\nonumber\\ 
&&\sum_{k=1}^n\,(-1)^{i+k}\, N(x_i,y_k)\,
D_{n-1}\left(\begin{array}{cccccc}
x_1 & \dots & \notxi & \dots & \dots & x_n \\
{}&{}&{}&{}&{}&{}\\ y_1 & \dots & \dots & \notyk & \dots & y_n 
\end{array}\Bigg|\lambda\right)\nonumber\\{}\nonumber\\
&&+\lambda
\int\limits_\Om N(x_i,s)\,D_n\left(\begin{array}{ccccc}
x_1 & \dots & (\!\!\notxi)s & \dots & x_n \\
{}&{}&{}&{}&{}\\ y_1 &  \dots & \dots & \dots & y_n 
\end{array}\Bigg|\lambda\right)\,ds 
\eeqra
where the symbol $\notxi$ in the upper row of $D_{n-1}$ indicates that 
the row index $x_i$ is to be omitted from the string $x_1,\ldots, x_n$ (and 
similarly for $\notyk$ in the lower row there), and $(\!\!\notxi)s$ indicates 
that $x_i$ in the upper row of $D_n$ under the integral should be replaced 
by the integration variable $s$. 

The second integral equation satisfied by $D_n$ is obtained similarly, by 
expanding each of the determinants in the series (\ref{nminor}) with respect 
to the column $y_i$. One obtains 
\beqra\label{yicolumn}
&&D_n\left(\begin{array}{ccc} x_1 & \dots & x_n \\
{}&{}&{}\\ y_1 & \dots & y_n 
\end{array}\Bigg|\lambda\right) =\nonumber\\{}\nonumber\\ 
&&\sum_{k=1}^n\,(-1)^{i+k}\, N(x_k,y_i)\,
D_{n-1}\left(\begin{array}{cccccc}
x_1 & \dots & \notxk & \dots & \dots & x_n \\
{}&{}&{}&{}&{}&{}\\ y_1 & \dots & \dots & \notyi & \dots & y_n 
\end{array}\Bigg|\lambda\right)\nonumber\\{}\nonumber\\
&&+\lambda
\int\limits_\Om N(s,y_i)\,D_n\left(\begin{array}{ccccc}
x_1 & \dots & \ldots & \dots & x_n \\
{}&{}&{}&{}&{}\\ y_1 &  \dots & (\!\!\notyi)s & \dots & y_n 
\end{array}\Bigg|\lambda\right)\,ds 
\eeqra
We now proceed to derive our own results. We assume henceforth that 
$D(\lambda)\neq 0$. In this case, according to (\ref{uniquesolution}), 
Fredholm's equation (\ref{fredholm}) has a unique
solution
\beq\label{uniquesolutiontext}
\phi(x) = f(x) + \lambda\int\limits_\Om  R(x,y;\lambda)f(y)\,dy\,.
\eeq
From this solution we construct the quantity 
\beq\label{Chi}
\Xi_n\left(\begin{array}{ccccc}
x_1 & \dots & \dots & \dots & x_n \\
{}&{}&{}&{}&{}\\ y_1 &  \dots & \notyi & \dots & y_n 
\end{array}\Bigg|\lambda\right) = 
\lambda\int\limits_\Om \,D_n\left(\begin{array}{ccccc}
x_1 & \dots & \dots & \dots & x_n \\
{}&{}&{}&{}&{}\\ y_1 &  \dots & y_i & \dots & y_n 
\end{array}\Bigg|\lambda\right)\,\phi(y_i)\,dy_i 
\eeq
Then, expand the $D_n$ under the integral on the right-hand side of 
(\ref{Chi}) according to (\ref{yicolumn}). By exploiting the fact that 
$\lambda\int\limits_\Om\,N(x,y)\phi(y)\,dy = \phi(x) - f(x) = \lambda
\int\limits_\Om\, R(x,y;\lambda)\,f(y)\,dy$ from (\ref{fredholm}) and 
(\ref{uniquesolutiontext}), we note that there appears a $\Xi_n$ on the 
right-hand side of the equation, which cancels the original one on the left, 
leaving us with the identity
\beqra\label{fidentity}
&&\int\limits_\Om \,D_n\left(\begin{array}{ccccc}
x_1 & \dots & \dots & \dots & x_n \\
{}&{}&{}&{}&{}\\ y_1 &  \dots & y_i & \dots & y_n 
\end{array}\Bigg|\lambda\right)\,f(y_i)\,dy_i = \nonumber\\{}\nonumber\\
&&\sum_{k=1}^n\,(-1)^{i+k}\,\int\limits_\Om\, R(x_k,y_i;\lambda)\,
D_{n-1}\left(\begin{array}{cccccc}
x_1 & \dots & \notxk & \dots & \dots & x_n \\
{}&{}&{}&{}&{}&{}\\ y_1 & \dots & \dots & \notyi & \dots & y_n 
\end{array}\Bigg|\lambda\right)\,f(y_i)\,dy_i\,.\nonumber\\{}
\eeqra
Since this identity holds for all admissible given functions $f(x)$, we 
conclude that $D_n$ must satisfy the recursion relation 
\beq\label{yirecursionrelation}
D_n\left(\begin{array}{ccc} x_1 & \dots & x_n \\
{}&{}&{}\\ y_1 & \dots & y_n 
\end{array}\Bigg|\lambda\right) =
\sum_{k=1}^n\,(-1)^{i+k}\, R(x_k,y_i;\lambda)\,
D_{n-1}\left(\begin{array}{cccccc}
x_1 & \dots & \notxk & \dots & \dots & x_n \\
{}&{}&{}&{}&{}&{}\\ y_1 & \dots & \dots & \notyi & \dots & y_n 
\end{array}\Bigg|\lambda\right)
\eeq
or equivalently, 
\beq\label{yirecursionrelation1}
D(\lambda)D_n\left(\begin{array}{ccc} x_1 & \dots & x_n \\
{}&{}&{}\\ y_1 & \dots & y_n 
\end{array}\Bigg|\lambda\right) =
\sum_{k=1}^n\,(-1)^{i+k}\, D(x_k,y_i;\lambda)\,
D_{n-1}\left(\begin{array}{cccccc}
x_1 & \dots & \notxk & \dots & \dots & x_n \\
{}&{}&{}&{}&{}&{}\\ y_1 & \dots & \dots & \notyi & \dots & y_n 
\end{array}\Bigg|\lambda\right)\,,
\eeq
from (\ref{resolvent}). Similarly, from (\ref{xirow}), and by exploiting 
the associated (or transposed) Fredholm equation $\psi(x) = g(x) + 
\lambda\int\limits_\Om \psi(y)N(y,x)\,dy\,,$ we obtain 
the transposed identity 
\beq\label{xirecursionrelation}
D(\lambda)D_n\left(\begin{array}{ccc} x_1 & \dots & x_n \\
{}&{}&{}\\ y_1 & \dots & y_n 
\end{array}\Bigg|\lambda\right) =
\sum_{k=1}^n\,(-1)^{i+k}\, D(x_i,y_k;\lambda)\,
D_{n-1}\left(\begin{array}{cccccc}
x_1 & \dots & \notxi & \dots & \dots & x_n \\
{}&{}&{}&{}&{}&{}\\ y_1 & \dots & \dots & \notyk & \dots & y_n 
\end{array}\Bigg|\lambda\right)\,.
\eeq
The two identities (\ref{yirecursionrelation1}) and 
(\ref{xirecursionrelation}) strongly suggest that the quantity  
\beq\label{Deltan}
\Delta_n\left(\begin{array}{ccc} x_1 & \dots & x_n \\
{}&{}&{}\\ y_1 & \dots & y_n 
\end{array}\Bigg|\lambda\right) = {1\over D(\lambda)}
D_n\left(\begin{array}{ccc} x_1 & \dots & x_n \\
{}&{}&{}\\ y_1 & \dots & y_n 
\end{array}\Bigg|\lambda\right)
\eeq
is simply the $n\times n$ determinant with entries $R(x_i,y_j;\lambda)$ 
and corresponding minor 
$$\Delta_{n-1}\left(\begin{array}{cccccc}
x_1 & \dots & \notxk & \dots & \dots & x_n \\
{}&{}&{}&{}&{}&{}\\ y_1 & \dots & \dots & \notyi & \dots & y_n 
\end{array}\Bigg|\lambda\right)\,.$$
The proof of this proposition by induction is almost trivial: This proposition
is indeed the content of (\ref{yirecursionrelation1}) and 
(\ref{xirecursionrelation}) for $n=2$:
\beq\label{n2}
D(\lambda)D_2\left(\begin{array}{cc} x_1 & x_2 \\
{}&{}\\ y_1 & y_2 
\end{array}\Bigg|\lambda\right) =  D(x_1,y_1;\lambda) D(x_2,y_2;\lambda) -  
D(x_1,y_2;\lambda) D(x_2,y_1;\lambda)\,.
\eeq
(For $n=1$ (\ref{yirecursionrelation1}) and (\ref{xirecursionrelation}) 
yield a trivial identity.) Then, by assuming it holds for $\Delta_{n-1}$, we 
apply it to the $\Delta_{n-1}$\,s which appear on the right-hand sides of 
(\ref{yirecursionrelation1}) and (\ref{xirecursionrelation}), and thus 
observe that the latter are just the expansion of an $n\times n$ determinant
with entries $ R(x_i,y_j;\lambda)$ according to the $i$ column and $i$th row,
respectively. The proposition of the induction is thus verified for 
$\Delta_n$ as well.

Thus, we have derived our first main result:
\beq\label{determinantalrep}
{1\over D(\lambda)}
D_n\left(\begin{array}{ccc} x_1 & \dots & x_n \\
{}&{}&{}\\ y_1 & \dots & y_n 
\end{array}\Bigg|\lambda\right) = \det_{ij} R(x_i,y_j;\lambda)\,.
\eeq
Note that for $n=1$, (\ref{determinantalrep}) coincides with 
(\ref{resolvent}), as it should.

\section{Interpretation of (\ref{determinantalrep}) in Terms of Non-interacting
Fermions}
\setcounter{equation}{0}
\renewcommand{\theequation}{3.\arabic{equation}}

The determinantal representation (\ref{determinantalrep}) suggests,
due to Wick's theorem for non-interacting fermions, an interpretation of 
$D_n$ as the correlation function of $n$ fermions and $n$ anti-fermions.
  
To this end, consider the non-interacting complex Grassmann valued field 
$\psi(x)$, living on $\Om$, with action 
\beq\label{action}
S = \int\limits_\Om \psi^\dgg(x)\left[\delta(x-y) - \lambda N(x,y)\right]\,
dx dy\,.
\eeq
Its partition function\cite{grassmann} is given by the path integral 
\beq\label{partition} 
\cz = \int\,\cd \psi^\dgg\,\cd \psi\,e^S = 
{\rm Det}\left({\bf 1} - \lambda\hat N\right) = D(\lambda)\,.
\eeq
As is well-known from the annals of quantum field theory, the non-vanishing 
correlation functions of (\ref{partition}) are those which contain equal 
numbers of $\psi$\,s and $\psi^\dgg$\,s, namely,
\beq\label{correlationdef}
\langle\,\psi^\dgg (y_1)\psi(x_1)\cdots \psi^\dgg (y_n)\psi(x_n)\,\rangle = 
{1\over \cz} \int\,\cd \psi^\dgg\,\cd \psi\, e^S\,
\left[\psi^\dgg (y_1)\psi(x_1)\cdots \psi^\dgg (y_n)\psi(x_n)\right]\,.
\eeq
The latter are determined according to Wick's theorem as 
\beq\label{wick}
\langle\,\psi^\dgg (y_1)\psi(x_1)\cdots \psi^\dgg (y_n)\psi(x_n)\,\rangle = 
\det_{ij} \langle\,\psi^\dgg (y_i)\psi(x_j)\,\rangle\,, 
\eeq
where the two-point function is 
\beq\label{2point}
\langle\,\psi^\dgg (y)\psi(x)\,\rangle = \langle x\,|{1\over {\bf 1} - \lambda
\hat N}\,|y\rangle\,.
\eeq
Then, observe from (\ref{2point}) and (\ref{resolventop}) that 
\beq\label{r2point}
\int\limits_\Om N(x,z)\langle\,\psi^\dgg (y)\psi(z)\,\rangle \,dz
= \langle x\,|{\hat N\over {\bf 1} - \lambda
\hat N}\,|y\rangle = R(x,y;\lambda)\,.
\eeq
Thus, by linearity, we obtain from (\ref{wick}) and (\ref{r2point}) that 
\beq\label{rdet}
\langle\,\psi^\dgg (y_1)(\hat N\psi)(x_1)\cdots 
\psi^\dgg (y_n)(\hat N\psi)(x_n)\,\rangle = 
\det_{ij}  R(x_i,y_j;\lambda)\,,
\eeq
where $(\hat N\psi)(x) = \int\limits_\Om N(x,y)\psi(y)\,dy$. 
Thus, we interpret the $n$th minor $D_n$, according to 
(\ref{determinantalrep}) and (\ref{rdet}), as the multi-fermion 
correlator
\beqra\label{fermioninterpretation}
&&D_n\left(\begin{array}{ccc} x_1 & \dots & x_n \\
{}&{}&{}\\ y_1 & \dots & y_n 
\end{array}\Bigg|\lambda\right) = 
D(\lambda)\,\langle\,\psi^\dgg (y_1)(\hat N\psi)(x_1)\cdots 
\psi^\dgg (y_n)(\hat N\psi)(x_n)\,\rangle\nonumber\\{}\nonumber\\
&&=\int\,\cd \psi^\dgg\,\cd \psi\, e^S\,
\left[\psi^\dgg (y_1)(\hat N\psi)(x_1)\cdots 
\psi^\dgg (y_n)(\hat N\psi)(x_n)\right]\,,
\eeqra
that is, $D_n$ is the $2n$th moment of the Grassmann weight $e^S$ (convoluted
against $n$ powers of $\hat N$). As such, it might be thought of as some kind
of a {\em continuum} (supplementary) compound matrix associated with 
${\bf 1} - \lambda\hat N$ \cite{compound}. The latter interpretation 
might be useful in studying minors of very large order, such that the $2n$ 
points $x_i$ and $y_i$ become typically dense in $\Omega$.

As a consistency check of (\ref{fermioninterpretation}), let us trace it over
all coordinates, and see if we recover (\ref{nthderivative}). Thus, 
\beqra\label{tracefermion}
&&\int\limits_\Om\,D_n\left(\begin{array}{ccc} x_1 & \dots & x_n \\
{}&{}&{}\\ x_1 & \dots & x_n \end{array}\Bigg|\lambda\right)\,dx_1\ldots
dx_n  =\nonumber\\{}\nonumber\\
&& D(\lambda)\,\int\limits_\Om\,\langle\,\psi^\dgg (x_1)(\hat N\psi)(x_1)
\cdots 
\psi^\dgg (y_n)(\hat N\psi)(x_n)\,\rangle\,dx_1\ldots
dx_n  
= \nonumber\\{}\nonumber\\
&& D(\lambda)\,\Bigg\langle\,\left(\int\limits_\Om\,\psi^\dgg (x)(\hat N\psi)
(x)\,dx 
\right)^n\,\Bigg\rangle\,.
\eeqra
From (\ref{action}), (\ref{partition}) and (\ref{correlationdef}), we see 
that 
\beq\label{fermionconsistency}
\Bigg\langle\,\left(\int\limits_\Om\,\psi^\dgg (x)(\hat N\psi)(x)\right)^n
\,\Bigg\rangle = {1\over D(\lambda)}\left(-\frac{d}{d\lambda}\right)^n 
D(\lambda)\,.
\eeq
Thus,
$$\int\limits_\Om\,D_n\left(\begin{array}{ccc} x_1 & \dots & x_n \\
{}&{}&{}\\ x_1 & \dots & x_n \end{array}\Bigg|\lambda\right)\,dx_1\ldots
dx_n  = \left(-\frac{d}{d\lambda}\right)^n 
D(\lambda)\,,$$
in accordance with (\ref{nthderivative}). 

\newpage
\section{The Functional Derivative of $D_n$ with Respect to the Kernel}
\setcounter{equation}{0}
\renewcommand{\theequation}{4.\arabic{equation}}
As we have discussed above, the minors $D_n$ (\ref{nminor}) determine
the solution of the Fredholm equation (\ref{fredholm}). In some applications
of (\ref{fredholm}), the kernel $N(x,y)$ may depend on a set of parameters
or functions, and it may be important to determine how the solutions vary with 
these quantities. To this end we have first to determine the functional 
derivative of the minors $D_n$ with respect to the kernel.

For example, in a recent paper \cite{glm}, we have calculated the variation 
of the solution of the Gelfand-Levitan-Marchenko equation with the reflection 
amplitude of scattering theory, and deduced from it the corresponding 
variation of the Schr\"odinger potential and wave-function. 

The minor $D_n$ is expressed in (\ref{determinantalrep}) in terms 
of $D(\lambda)$ and $R(x,y;\lambda)$. It is straightforward to obtain the
functional derivatives of these two objects with respect to the kernel 
$N(x,y)$ directly from (\ref{fredholmdeterminant}) and (\ref{resolventop}). 
Thus, consider a perturbation $\hat N\rightarrow \hat N + \delta\hat N$. 
From (\ref{fredholmdeterminant}) we see that under this variation 
$\delta D(\lambda) = D(\lambda)\delta\log D(\lambda) = 
-\lambda D(\lambda)\rmtr\left(\frac{1}{{\bf 1}-\lambda\hat N}
\delta\hat N\right)$, from which we infer 
\beqra\label{deltadeterminant}
{\delta D(\lambda)\over \delta N(a,b)} &=& -\lambda D(\lambda)\left[
\delta(b-a) + \lambda R(b,a;\lambda)\right]\nonumber\\
&=& -\lambda D(\lambda)\delta(b-a) -\lambda^2 D(b,a;\lambda)\,.
\eeqra
Under this variation we also have 
$\hat R \rightarrow \hat R +  \frac{1}{{\bf 1}-\lambda\hat N} 
\delta\hat N \frac{1}{{\bf 1}-\lambda\hat N}$. Consequently 
\beqra\label{deltaresolvent}
{\delta R(x,y;\lambda)\over \delta N(a,b)} &=& 
\langle x | ({\bf 1}+\lambda\hat R) |a \rangle \langle b | 
 ({\bf 1}+\lambda\hat R) |y \rangle\nonumber\\
&=& \left(\delta(x-a) + \lambda R(x,a;\lambda)\right)
\, \left(\delta(b-y) + \lambda R(b,y;\lambda)\right)\,.
\eeqra
We can then calculate $${\delta\over\delta N(a,b)}\, 
\left(D(\lambda)\det_{ij}R(x_i,y_j;\lambda)\right)$$ by applying 
(\ref{deltadeterminant}) and (\ref{deltaresolvent}) as necessary. The 
expression we obtain in this way is rather cumbersome, but the plethora of 
terms thus obtained can be organized into a linear combination of the minors
$D_n$ and $D_{n\pm1}$.

Instead of pursuing this line of derivation, we shall now sketch the 
calculation of the functional derivative of $D_n$ directly 
from (\ref{nminor}), by taking the derivative of this series term by term. 
Thus, consider taking the derivative of the $p$th term. Let us split the 
matrix whose determinant 
\beq\label{integrateddet}
\int\limits_\Om\,N\left(\begin{array}{cccccccc} x_1 & x_2 & \dots & x_n &
 s_{n+1} & s_{n+2} & \dots & s_{n+p} \\
{}&{}&{}&{}&{}&{}&{}&{}\\ y_1 & y_2 & \dots & y_n &
 s_{n+1} & s_{n+2} & \dots & s_{n+p}  \end{array}\right)\,ds_{n+1}\ldots 
ds_{n+p}
\eeq 
is being integrated in that term into four blocks, according to NN, NI, IN 
and II, where `I' stands for an integrated coordinate index, and `N' for 
a non-integrated one. Let us now scan systematically through these blocks.

When the derivative ${\delta\over\delta N(a,b)}$ hits the term $N(x_i,y_j)$ 
in the NN sector, it produces a factor ${\delta N(x_i,y_j)\over\delta N(a,b)} 
= \delta (x_i-a)\delta (y_j-b)$ which 
is multiplied by the minor of $N(x_i,y_j)$, times a sign factor $(-1)^{i+j}$.
The total contribution of the NN sector to the derivative is the sum of 
all these terms:
\beqra\label{NN}
{\rm NN} &=&\sum_{i,j =1}^n\,(-1)^{i+j}\, \delta (x_i-a)\delta (y_j-b)\,\times
\nonumber\\{}\nonumber\\
&&\int\limits_\Om\,N\left(\begin{array}{ccccccccc} x_1 & \dots & \notxi & 
\dots & \dots & x_n & s_{n+1} & \dots & s_{n+p} \\
{}&{}&{}&{}&{}&{}&{}&{}&{}\\ y_1 & \dots & \dots & \notyj & \dots & y_n &
 s_{n+1} & \dots & s_{n+p}  \end{array}\right)\,ds_{n+1}\ldots 
ds_{n+p}\,.\nonumber\\{}
\eeqra
Move now to the NI block. When the derivative ${\delta\over\delta N(a,b)}$ 
hits the term $N(x_i,s_{n+l})$, it produces a factor 
$\delta (x_i-a)\delta (s_{n+l}-b)$
which is multiplied by the minor of $N(x_i,s_{n+l})$, which contains 
$s_{n+l}$ as a row index (but not as a column index), times a sign factor 
$(-1)^{i+n+l}$. Integration over $s_{n+l}$ thus replaces the row index
$s_{n+l}$ in that minor by $b$. Now, permute the row which used to be 
that of $s_{n+l}$ in that minor, and move it in between the rows corresponding 
to $x_{i-1}$ and $x_{i+1}$. This means permuting it across $n+l-i-1$ rows and 
costs a sign factor $(-1)^{n+l-i-1}$, which combines with the previous sign
factor to $(-1)$, independently of $l$. Thus, all the $p$ columns which 
intersect the original row $x_i$ in the NI block make the same contribution
to the functional derivative, after integration over the remaining $p-1$ 
variables. Finally, summing over all the $x_i$ in the NI block we obtain
the total contribution by that block to the functional derivative as 
\beqra\label{NI}
&&{\rm NI} =-p\sum_{i=1}^n\,\delta (x_i-a)\,\times
\nonumber\\
&&\int\limits_\Om\,N\left(\begin{array}{cccccccc} x_1 & \dots & 
(\!\!\notxi)b & 
\dots & x_n & s_{n+1} & \dots & s_{n+p-1} \\
{}&{}&{}&{}&{}&{}&{}&{}\\ y_1 & \dots & \dots & \dots & y_n &
 s_{n+1} & \dots & s_{n+p-1}  \end{array}\right)\,ds_{n+1}\ldots 
ds_{n+p-1}\,.{}
\eeqra
Similarly, the contribution of the entire IN block to the functional 
derivative is 
\beqra\label{IN}
{\rm IN} &=&-p\sum_{j=1}^n\,\delta (y_j-b)\,\times
\nonumber\\{}\nonumber\\
&&\int\limits_\Om\,N\left(\begin{array}{cccccccc} x_1 & \dots & 
\dots & \dots & x_n & s_{n+1} & \dots & s_{n+p-1} \\
{}&{}&{}&{}&{}&{}&{}&{}\\ y_1 & \dots & (\!\!\notyj)a & \dots & y_n &
 s_{n+1} & \dots & s_{n+p-1}  \end{array}\right)\,ds_{n+1}\ldots 
ds_{n+p-1}\,.\nonumber\\{}
\eeqra

The last block is II. Clearly, we should discuss the diagonal terms and the 
non-diagonal terms separately. 
When ${\delta\over\delta N(a,b)}$ hits the diagonal term $N(s_{n+l},s_{n+l})$,
it produces a factor $\delta (s_{n+l}-a)\delta (s_{n+l}-b)$
times the corresponding diagonal minor, which comes with a positive sign 
and does not depend on $s_{n+l}$. Thus, integration over $s_{n+l}$ simply 
produces a factor $\delta (a-b)$ which multiplies the remaining integral. 
The latter is the same for all diagonal terms. Thus, the overall contribution
of diagonal terms from the II block to the functional derivative is
\beqra\label{IIdiag}
{\rm II}_{diag} = 
p\,\delta (a-b)\,\int\limits_\Om\,N\left(\begin{array}{cccccc} 
x_1 & \dots & x_n & s_{n+1} & \dots & s_{n+p-1} \\
{}&{}&{}&{}&{}&{}\\ 
y_1 & \dots & y_n & s_{n+1} & \dots & s_{n+p-1}  
\end{array}\right)\,ds_{n+1}\ldots 
ds_{n+p-1}\,.\nonumber\\{}
\eeqra
Finally, when ${\delta\over\delta N(a,b)}$ hits the non-diagonal term 
$N(s_{n+l},s_{n+k})$, $l\neq k$, it produces a factor 
$\delta (s_{n+l}-a)\delta (s_{n+k}-b)$ times the minor of 
$N(s_{n+l},s_{n+k})$, which contains $s_{n+k}$ as a row index 
(but not as a column index), and also contains $s_{n+l}$ as a column index 
(but not as a row index), times a sign factor $(-1)^{n+k+n+l}$.

Integration over $s_{n+l}$ thus replaces the column index $s_{n+l}$ in that 
minor by $a$. Similarly, integration over $s_{n+k}$ replaces the row index 
$s_{n+k}$ by $b$. Now, move the row which used to be that of $s_{n+k}$ right 
below the row $x_n$, and the column which used to be that of $s_{n+l}$ 
immediately to the right of the column $y_n$. These
permutations produce a sign factor $(-1)^{k-1 + l-1 -1}$, which combines with
the previous sign simply to $(-1)$. The remaining integral is independent
of $s_k$ and $s_l$ and yields the same contribution for all the 
$p(p-1)$ non-diagonal terms. Thus, their total contribution to the 
functional derivative is 
\beqra\label{IInondiag}
{\rm II}_{non-diag.} =-p(p-1)\,\int\limits_\Om\,N\left(
\begin{array}{ccccccc} 
x_1 & \dots & x_n & b & s_{n+1} & \dots & s_{n+p-2} \\
{}&{}&{}&{}&{}&{}&{}\\ 
y_1 & \dots & y_n & a & s_{n+1} & \dots & s_{n+p-2}  
\end{array}\right)\,ds_{n+1}\ldots 
ds_{n+p-2}\,.\nonumber\\{}
\eeqra
Gathering all contributions (\ref{NN})-(\ref{IInondiag}) together, multiplying
their sum by ${(-\lambda)^p\over p!}$ and summing over $p$, we finally arrive,
after some rearrangement of terms, at our second main result:
\beqra\label{derivative}
&&{\delta\over\delta N(a,b)}\,
D_n\left(\begin{array}{ccc} x_1 & \dots & x_n \\
{}&{}&{}\\ y_1 & \dots & y_n 
\end{array}\Bigg|\lambda\right) =\nonumber\\{}\nonumber\\{}\nonumber\\
&&\sum_{i,j =1}^n\,(-1)^{i+j}\,\delta(x_i-a)\,\delta(y_j-b)
\,D_{n-1}\left(\begin{array}{cccccc}
x_1 & \dots & \notxi & \dots & \dots & x_n \\
{}&{}&{}&{}&{}&{}\\ y_1 & \dots & \dots & \notyj & \dots & y_n 
\end{array}\Bigg|\lambda\right)\nonumber\\{}\nonumber\\{}\nonumber\\
&&+\lambda\sum_{i=1}^n\,\delta(x_i-a)\,D_n\left(\begin{array}{ccccc}
x_1 & \dots & (\!\!\notxi)b & \dots & x_n \\
{}&{}&{}&{}&{}\\ y_1 &  \dots & \dots & \dots & y_n 
\end{array}\Bigg|\lambda\right)\nonumber\\{}\nonumber\\{}\nonumber\\
&&+\lambda\sum_{j=1}^n\,\delta(y_j-b)\,
D_n\left(\begin{array}{ccccc}
x_1 & \dots & \ldots & \dots & x_n \\
{}&{}&{}&{}&{}\\ y_1 &  \dots & (\!\!\notyj)a & \dots & y_n 
\end{array}\Bigg|\lambda\right)\nonumber\\{}\nonumber\\{}\nonumber\\
&&-\lambda\delta(a-b)\,D_n\left(\begin{array}{ccc} x_1 & \dots & x_n \\
{}&{}&{}\\ y_1 & \dots & y_n 
\end{array}\Bigg|\lambda\right)
-\lambda^2\,D_{n+1}\left(\begin{array}{cccc} x_1 & \dots & x_n & b\\
{}&{}&{}&{}\\ y_1 & \dots & y_n & a \end{array}\Bigg|\lambda\right)
\eeqra
It is gratifying that the functional derivative of $D_n$ is expressed
as a relatively simple linear combination of $D_n$ and $D_{n\pm 1}$. 
In particular, we could think of (\ref{derivative}) as recursively defining
$D_{n+1}$ in terms of the lower minors. This is analogous to Jacobi's 
recursive definition of supplementary compound matrices in the finite 
dimensional case \cite{compound}. The formuala (\ref{derivative}) for 
the functional derivative of $D_n$ should coincide, of course, with the 
expression we would obtain by taking the derivative of 
$D(\lambda)\det_{ij}R(x_i,y_j;\lambda)$. 

As a simple application, let us check (\ref{derivative}) for $n=0$ and 
$n=1$. For $n=0$, it yields
$${\delta D(\lambda)\over \delta  N(a,b)} = -\lambda\delta(a-b)\,D(\lambda)
-\lambda^2\,D (b,a;\lambda)\,,$$
which coincides with (\ref{deltadeterminant}). Similarly, for $n=1$, we 
obtain 
\beqra\label{nonecheck}
{\delta D(x,y;\lambda)\over \delta  N(a,b)} &=& 
\delta(x-a)\,\delta(y-b) D(\lambda) + \lambda\delta(x-a)\,D(b,y;\lambda)
+\lambda\delta(y-b)\,D(x,a;\lambda)\nonumber\\{}\nonumber\\
&-&\lambda\delta(a-b)\,D(x,y;\lambda)
-\lambda^2\,D_2
\left(\begin{array}{cc} x & b\\ y & a \end{array}\Bigg|\lambda\right)\,.
\eeqra
Thus, from the last two equation we can derive that  
\beqra\label{resolventcheck}
&&{\delta R(x,y;\lambda)\over \delta  N(a,b)} = 
{\delta \over \delta  N(a,b)}\left({ D(x,y;\lambda)\over  D(\lambda)}\right)=
\nonumber\\{}\nonumber\\
&&\delta(x-a)\,\delta(y-b) + \lambda\delta(x-a)\,R(b,y;\lambda)
+\lambda\delta(y-b)\,R(x,a;\lambda)\nonumber\\{}\nonumber\\
&&+\lambda^2\,\left( R(x,y;\lambda) R(b,a;\lambda)-
\Delta_2
\left(\begin{array}{cc} x & b\\ y & a \end{array}\Bigg|\lambda\right)\right)
\eeqra
which coincides with (\ref{deltaresolvent}) due to our determinantal 
representation (\ref{determinantalrep}).


\setcounter{equation}{0}
\setcounter{section}{0}
\renewcommand{\theequation}{A.\arabic{equation}}
\renewcommand{\thesection}{Appendix:}
\section{The Solution of Fredholm's Equation}
\vskip 5mm
\setcounter{section}{0}
\renewcommand{\thesection}{A}

The theory of the Fredholm's equation (\ref{fredholm}) and its solution 
is summarized in Fredholm's celebrated three theorems 
(sometimes referred to collectively as ``Fredholm's alternatives''). 
The exposition in this appendix will be very telegraphic. We will describe 
the content of Fredholm's theorems in a semi-quantitative way, sufficient for 
our purposes, and refer the interested reader to the cited literature for 
more details.

Broadly speaking, the solution of (\ref{fredholm}) depends on whether 
$D(\lambda)\neq 0$ or not:

~~~~~{\em Case (1):} $D(\lambda)\neq 0$ \\
In this case, the solution of (\ref{fredholm}) involves the $n=1$ minor 
$D (x,y; \lambda)$. For values of $\lambda$ such that $D(\lambda)\neq 0$, 
the operator ${\bf 1} -\lambda\hat N$ is invertible, and (\ref{fredholm}) 
(or, equivalently (\ref{fredholmop})) has a {\em unique} solution, given by 
\beq\label{uniquesolution}
\phi(x) = f(x) + \lambda\int\limits_\Om  R(x,y;\lambda)f(y)\,dy\,,
\eeq
where  $R(x,y;\lambda)$ is the resolvent kernel of (\ref{fredholm}), which
is given by 
\beq\label{resolvent}
R(x,y;\lambda) = {D (x,y; \lambda)\over D(\lambda)}\,.
\eeq
From (\ref{fredholmop}) we can read off the operator $\hat R$, which
corresponds to $R(x,y;\lambda) = \langle x | \hat R  |y \rangle$  as 
\beq\label{resolventop}
\hat R = {1\over\lambda}
\left(\frac{1}{{\bf 1}-\lambda\hat N} - {\bf 1}\right) = 
\frac{\hat N}{1- \lambda\hat N}\,. 
\eeq
Thus, from (\ref{resolvent}), (\ref{nthderivative}) (at $n=1$), 
(\ref{fredholmdeterminant}) and (\ref{resolventop}) we conclude that
\beq\label{traceresolvent}
\rmtr\hat R = \int\limits_\Om  R(x,x;\lambda)\,dx = -\frac{d}{d\lambda}
\log\,D(\lambda)\,,
\eeq
which shows that the poles of $\hat R$ as a function of $\lambda$ are the 
zeros of $D(\lambda)\,.$

~~~~~{\em Case (2):} $D(\lambda)= 0$ \\
If $\lambda=\lambda_0$ such that $D(\lambda_0)= 0$, the homogeneous 
equation 
\beq\label{homogeneous}
\phi(x) = \lambda_0\int\limits_\Om N(x,y)\phi(y)\,dy
\eeq
has one or more non-trivial, linearly independent solutions $\Phi_i(x)\,,
\quad i=1,\ldots \nu$, where $\nu\geq 1 $. In this case, we say that 
$\lambda_0$ is an eigenvalue of $\hat N$ of rank $\nu$, and refer
to the functions $\Phi_i(x)$ as the characteristic functions corresponding to 
the eigenvalue $\lambda_0$. (This nomenclature deviates from that of linear 
algebra, which would refer to $\frac{1}{\lambda_0}$ as the eigenvalue.) 
Any solution of the homogeneous equation (\ref{homogeneous}) is a linear
combination of the characteristic functions, i.e., the characteristic 
functions span ${\rm Ker}(1- \lambda_0\hat N)$.

As it turns out, the $\nu$ characteristic functions are proportional to 
the $\nu$th minor $D_\nu$. More precisely, for a fixed set of $2\nu$ points 
$x_1,\ldots y_\nu$, such that 
$$D_\nu\left(\begin{array}{ccc} x_1 & \dots & x_\nu \\
{}&{}&{}\\ y_1 & \dots & y_\nu 
\end{array}\Bigg|\lambda_0\right)\neq 0\,,$$ we have 
\beq\label{Phii}
\Phi_i(x) = {D_\nu\left(\begin{array}{ccccc}
x_1 & \dots & (\!\!\notxi)x & \dots & x_\nu \\
{}&{}&{}&{}&{}\\ y_1 &  \dots & y_i & \dots & y_\nu 
\end{array}\Bigg|\lambda_0\right)\over
D_\nu\left(\begin{array}{ccc} x_1 & \dots & x_\nu \\
{}&{}&{}\\ y_1 & \dots & y_\nu 
\end{array}\Bigg|\lambda_0\right)}\,,
\eeq
where the symbol $(\!\!\notxi)x$ indicates that the $i$th row index 
$x_i$ in the upper row of the minor in the numerator is to be replaced by 
the coordinate $x$. The functions (\ref{Phii}) are normalized such that 
\beq\label{Phinormalization}
\Phi_i(x_k) = \delta_{i,k}\,,
\eeq
as can be seen from (\ref{nminor}), since $D_n$ vanishes when any two of its 
row (or column) indices coincide.

Since the characteristic functions are expressed in terms of
$D_\nu$, it does not vanish identically. In fact, it is the minor of lowest 
order which does not vanish identically as a function of its $2\nu$ 
arguments at $\lambda = \lambda_0$. In fact, (\ref{Phii}) is obtained by 
setting $n=\nu$ (and $\lambda=\lambda_0$) in (\ref{xirow}). Furthermore, it 
follows from (\ref{nthderivative}) that $\lambda_0$ must be a zero of 
$D(\lambda)$ of multiplicity which is greater or equal to $\nu$, since $
\frac{d^\nu D(\lambda)}{d\lambda^\nu}$ might still vanish at 
$\lambda=\lambda_0$. 

As for solving the inhomogeneous equation (\ref{fredholm}) at 
$\lambda=\lambda_0$, one proceeds as follows. First, one has to consider the 
transposed, or associated homogeneous Fredholm equation 
\beq\label{associated} 
\psi(x) = \lambda_0\int\limits_\Om \psi(y)N(y,x)\,dy\,.
\eeq
Since its kernel is the transpose of the kernel of (\ref{fredholm}),
it has $\lambda_0$ as an eigenvalue of the same rank $\nu$. Thus, 
there are $\nu$ independent characteristic solutions $\Psi_i(x)$,
which are also expressed in terms of the $\nu$th minor $D_\nu$, similarly to 
(\ref{Phii}), as 
\beq\label{Psii}
\Psi_i(x) = {D_\nu\left(\begin{array}{ccccc}
x_1 & \dots & x_i & \dots & x_\nu \\
{}&{}&{}&{}&{}\\ y_1 &  \dots & (\!\!\notyi)x & \dots & y_\nu 
\end{array}\Bigg|\lambda_0\right)\over
D_\nu\left(\begin{array}{ccc} x_1 & \dots & x_\nu \\
{}&{}&{}\\ y_1 & \dots & y_\nu 
\end{array}\Bigg|\lambda_0\right)}\,,
\eeq
which span  ${\rm Ker}(1- \lambda_0\hat N^T)$. 

A necessary and sufficient condition for the existence of a solution of 
(\ref{fredholm}) at $\lambda=\lambda_0$ is then that 
the given function $f(x)$ be orthogonal to all the characteristic functions
$\Psi_i(x)$, i.e., that $\int\limits_\Om \Psi_i(x) f(x) =0\,,
\quad i=1\,,\ldots\,,\nu$.

If this condition holds, the solution (which exists) is not unique, since 
given a particular solution, one can always add to it an arbitrary solution of 
the homogeneous equation (\ref{homogeneous}). The piece in the general 
solution of (\ref{fredholm}) which is linear in $f(x)$ (i.e., a particular
solution of (\ref{fredholm})) is 
\beq\label{particular}
\phi_p(x) = f(x) + \lambda_0\int\limits_\Om\,{D_{\nu+1}\left(\begin{array}
{cccc} x & x_1 & \dots & x_\nu \\
{}&{}&{}&{}\\ y & y_1 &  \dots & y_\nu 
\end{array}\Bigg|\lambda_0\right)\over
D_\nu\left(\begin{array}{ccc} x_1 & \dots & x_\nu \\
{}&{}&{}\\ y_1 & \dots & y_\nu 
\end{array}\Bigg|\lambda_0\right)}\,f(y)\,dy\,.
\eeq

\pagebreak

\section*{Acknowledgments}
I thank Ady Mann for making some useful comments. I also 
thank Klaus Scharnhorst for pointing out the analogy with 
compound matrices in the finite dimensional case and for turning my attention 
to the papers in \cite{compound}.

\end{document}